# A Compact Exponential Scheme for Solving 1D Unsteady Convection-Diffusion Equation with Neumann Boundary Conditions

**Yucheng Fu[a], Zhenfu Tian*[, b], Yang Liu[a]**

[a]Nuclear Engineering Program, Mechanical Engineering Department, Virginia Tech
635 Prices Fork Road, Blacksburg, VA 24061

[b]Department of Mechanics and Engineering Science, Fudan University
Shanghai 200433, P.R. China

ycfu@vt.edu; zftian@fudan.edu.cn; liu130@vt.edu

## Abstract

In this paper, a high-order exponential scheme is developed to solve the 1D unsteady convection-diffusion equation with Neumann boundary conditions. The present method applies fourth-order compact exponential difference scheme in spatial discretization at all interior and boundary points. The Padé approximation is used for the discretization. The resulting scheme obtains fourth-order accuracy in both spatial and temporal discretization. In each iterative loop, the scheme corresponds to a strictly diagonally dominant tridiagonal matrix equation, which can be inverted by simple tridiagonal Gaussian decomposition. The developed scheme is proved numerically unconditionally stable for convection dominated problems. Four typical PDEs with Neumann boundary conditions are provided to verify the accuracy of the proposed scheme. The results are compared with analytical solutions and numerical results calculated by different numerical methods. It shows that the new scheme produces high accuracy solutions for all the test problems and it is more suitable for dealing with convection dominated problems.

## Keywords

High-order exponential scheme, Neumann boundary, convection-diffusion equation, convection dominated, Padé approximation

## 1 Introduction

The convection-diffusion equation has wide applications, especially for those involving heat transfer and fluid flows. Examples of using convection-diffusion equation model to solve practical engineering problems include heat transfer problems in a draining film [1] or a nanofluid filled enclosure [2], radial transport in a porous medium [3], water transport in soils [4], *etc*. For convection-diffusion problems, only few cases with special initial conditions have analytical solutions. Therefore, development of stable, accurate and efficient numerical methods for solving convection-diffusion equations is of vital importance. In this paper, we mainly consider the following 1D unsteady convection-diffusion equation:

$$\frac{\partial u}{\partial t}+p\frac{\partial u}{\partial x}=a\frac{\partial^2 u}{\partial x^2},(x,t)\in[0,L]\times[0,T], \tag{1}$$

with initial value:

$$u(x,0)=\varphi(x), x\in[0,L], \tag{2}$$

and Neumann boundary condition:

$$\frac{\partial u(0,t)}{\partial x} = g_1(t), \frac{\partial u(L,t)}{\partial x} = g_2(t), t \in [0,T], \tag{3}$$

where $u(x,t)$, $\varphi(x)$, $g(x)$ are given sufficiently smooth functions. The term $u(x,t)$ can represent vorticity, temperature, mass concentration, or other physical quantities that are transferred inside the physical system by convection and diffusion. In Eq. (1), $p$ is the convection velocity and $a$ is the diffusivity, which has a positive value.

Many finite difference methods have been developed to solve Eq. (1). In recent years, the compact finite difference schemes have attracted more attention for its high spatial resolution and its compact stencil. Most of these schemes are proposed with Dirichlet boundary conditions [5–14]. For Neumann boundary, the development of high order compact scheme is more complicated. Chawla *et al*. [15] developed an extended one-step time-integration scheme for convection-diffusion equations. It is second-order on the Neumann boundary. To achieve high accuracy, some authors try to use high-order spatial derivatives to approximate the Neumann boundary conditions. Spotz and Carey [16] proposed a method to a achieve fourth-order accuracy for the unsteady convection-diffusion equations with Neumann boundary conditions. This scheme leads to a noncompact approximation for time steps. Zhao [17] constructed a fourth-order compact finite difference scheme for a heat conduction problem with Neumann boundary conditions. It has a fourth-order accuracy at the boundary points and has an overall accuracy of the order of $O(\Delta t^2 + h^4)$. Cao et al. [18] developed a scheme using a polynomial compact difference scheme for spatial variable and Padé approximation for temporal variable to solve the unsteady convection-diffusion equations. However, the fourth-order polynomial compact scheme is a lower resolution scheme. As shown in literature [19,20], it may not achieve a good resolution for solving the convection dominated problems.

In this work, we first introduce a three point $O(h^4)$ compact exponential finite difference scheme for 1D unsteady convection-diffusion equation by treating the time derivative as the function of $x$. To achieve a global fourth-order accuracy in space, we also propose a fourth-order exponential difference scheme in the boundary points. In Section 3, the stability of the present high order exponential scheme is analyzed. Numerical experiments are carried out in Section 4 to verify the accuracy of the proposed scheme. Concluding remarks and forecasts are included in Section 5. A detailed proof of strictly diagonal dominance of the matrix is provided in Appendix B.

## 2 Development of the fourth-order exponential scheme

### 2.1 The interior spatial points

To develop the finite difference scheme, the interested domain is discretized into grid points for numerical calculation. These grid points are labeled sequentially in space as $x_0, x_1, \ldots, x_N$ and in time as $t_0, t_1, \ldots, t_M$. For a uniform grid with space step size $h$ and time step size $\Delta t$, the discrete grid points and times within a given spatial domain $[0,l]$ and time domain $[0,T]$ are calculated as:

$$x_i = ih,\ i = 0,1,\ldots,N, \tag{4}$$

$$t_n = n\Delta t,\ n = 0,1,\ldots,M, \tag{5}$$

where $h = l/N$ and $\Delta t = T/M$. The quantity $u(x_i,t_n)$ represents the exact solution at $(x_i,t_n)$. The term $u_i^n$ is used to represents the numerical solution at $(x_i,t_n)$.

Consider the 1D steady-state convection-diffusion equation first:

$$-a\frac{\partial^2 u}{\partial x^2}+p\frac{\partial u}{\partial x}=f, \tag{6}$$

where $a$ is the constant diffusivity, $p$ is the constant convective velocity, $f$ is a sufficiently smooth function of $x$ and $u$, which may represent vorticity, speed, etc. For the interior points, a three-point fourth-order compact finite difference formulation can be given symbolically as [13,21] (see Appendix A):

$$-\alpha\delta_x^2 u_i+p\delta_x u_i=f_i+\alpha_1\delta_x f_i+\alpha_2\delta_x^2 f_i, \tag{7}$$

where the first order and second order central difference operators $\delta_x u_i$ and $\delta_x^2 u_i$ are defined respectively, as:

$$\delta_x u_i=\frac{u_{i+1}-u_{i-1}}{2h},\ \text{and}\ \delta_x^2 u_i=\frac{u_{i+1}-2u_i+u_{i-1}}{h^2}. \tag{8}$$

The parameters $\alpha,\alpha_1$ and $\alpha_2$ are defined as:

$$\alpha=\begin{cases}\frac{ph}{2}\coth\left(\frac{ph}{2a}\right), & p\neq 0,\\ a, & p=0,\end{cases}\quad \alpha_1=\begin{cases}\frac{a-\alpha}{p}, & p\neq 0,\\ 0, & p=0,\end{cases}\quad \alpha_2=\begin{cases}\frac{a(a-\alpha)}{p^2}+\frac{h^2}{6}, & p\neq 0,\\ \frac{h^2}{12}, & p=0.\end{cases} \tag{9}$$

It is easy to find that Eq.(7) consists of a diagonally dominant tri-diagonal system of equations. Substitute $f$ in Eq. (7) by $-\left(\frac{\partial u}{\partial t}\right)_i$, a semi-discrete fourth-order exponential approximation for Eq. (1) on inner spatial grids is developed as:

$$\begin{aligned}\left(\frac{\alpha_2}{h^2}-\frac{\alpha_1}{2h}\right)\left(\frac{\partial u}{\partial t}\right)_{i-1}^n+\left(1-\frac{2\alpha_2}{h^2}\right)\left(\frac{\partial u}{\partial t}\right)_i^n+\left(\frac{\alpha_2}{h^2}+\frac{\alpha_1}{2h}\right)\left(\frac{\partial u}{\partial t}\right)_{i+1}^n=\\ \left(\frac{\alpha}{h^2}+\frac{p}{2h}\right)u_{i-1}^n-\frac{2\alpha}{h}u_i^n+\left(\frac{\alpha}{h^2}-\frac{p}{2h}\right)u_{i+1}^n.\end{aligned} \tag{10}$$

## 2.2 The boundary spatial points

The term $u(x,t)$ is assumed to be a sufficiently smooth function of $x$ and $t$. The derivative boundary conditions can be approximated by the following formulas:

$$\frac{\partial u(x_0,t)}{\partial x}=\frac{u(x_1,t)-u(x_0,t)}{h}+S_1h\frac{\partial^2 u(x_0,t)}{\partial x^2}+S_2h\frac{\partial^2 u(x_1,t)}{\partial x^2}+S_3h\frac{\partial^3 u(x_0,t)}{\partial x^3}+O(h^4), \tag{11}$$

and

$$\frac{\partial u(x_N,t)}{\partial x}=\frac{u(x_N,t)-u(x_{N-1},t)}{h}+S_4h\frac{\partial^2 u(x_N,t)}{\partial x^2}+S_5h\frac{\partial^2 u(x_{N-1},t)}{\partial x^2}+S_6h\frac{\partial^3 u(x_N,t)}{\partial x^3}+O(h^4). \tag{12}$$

In order to determine the parameters $S_1,...,S_6$, we require that the above formulas have the local exactness of the type $\left\{1,x,x^2,x^3,e^{\frac{px}{a}}\right\}$. Substituting $u(x,t)$ with these functions, the parameters $S_1,...,S_6$ are obtained by solving the system of equations:

$$S_1 = -\frac{1}{2} + \beta_1, S_2 = -\beta_1, S_3 = -\frac{1}{6} + \beta_1,$$
$$S_4 = \frac{1}{2} - \beta_2, \;\; S_5 = \beta_2, \;\; S_6 = \frac{1}{6} - \beta_2, \tag{13}$$

where,

$$\beta_1 = \begin{cases} \dfrac{a^2}{p^2h^2} + \dfrac{3 + \frac{ph}{a}}{6\left(1 - e^{\frac{ph}{a}} + \frac{ph}{a}\right)}, p \neq 0, \\ \dfrac{1}{12}, \qquad\qquad\qquad p = 0, \end{cases} \qquad \beta_2 = \begin{cases} \dfrac{a^2}{p^2h^2} + \dfrac{3 - \frac{ph}{a}}{6\left(1 - e^{-\frac{ph}{a}} - \frac{ph}{a}\right)}, p \neq 0, \\ \dfrac{1}{12}, \qquad\qquad\qquad p = 0. \end{cases} \tag{14}$$

Using Taylor series expansion, the coefficient $\beta_1$ can be approximated as:

$$\beta_1 = \frac{1 + \frac{1}{5}\left(\frac{ph}{a}\right) + \frac{1}{30}\left(\frac{ph}{a}\right) + o\left(h^3\right)}{12\left[1 + \frac{1}{3}\left(\frac{ph}{a}\right) + \frac{1}{12}\left(\frac{ph}{a}\right)^2 + o\left(h^3\right)\right]} = \frac{1}{12} - \frac{2}{15}\left(\frac{ph}{a}\right) + o\left(h^2\right). \tag{15}$$

Replace $\beta_1$ in Eq. (11) by Eq. (15), one has:

$$\frac{\partial u\left(x_0,t\right)}{\partial x} = \frac{\partial u\left(x_0,t\right)}{\partial x} + \frac{ph^4}{15a}\frac{\partial^4 u\left(x_0,t\right)}{\partial x^4} - \left(\frac{h^4}{180} - \frac{ph^5}{45a}\right)\frac{\partial^5 u\left(x_0,t\right)}{\partial x^5} + o\left(h^5\right). \tag{16}$$

This shows that the new proposed scheme achieves fourth-order accuracy for the Neumann boundary condition. The deduction of $\beta_2$ follows the same procedures.

To obtain the compact difference schemes on the boundary, the higher order derivatives in Eq. (11) and Eq. (12) should be replaced. From Eq. (1), we have

$$\frac{\partial^2 u\left(x_0,t\right)}{\partial x^2} = \frac{1}{a}\left(p\frac{\partial u\left(x_0,t\right)}{\partial x} + \frac{\partial u\left(x_0,t\right)}{\partial t}\right) = \frac{p}{a}g_1\left(T\right) + \frac{1}{a}\left(\frac{\partial u}{\partial t}\right)_0, \tag{17}$$

$$\begin{aligned} \frac{\partial^3 u\left(x_0,t\right)}{\partial x^3} &= \frac{p}{a^2}\left(p\frac{\partial u\left(x_0,t\right)}{\partial x} + \frac{\partial u\left(x_0,t\right)}{\partial t}\right) + \frac{1}{a}\frac{\partial^2 u\left(x_0,t\right)}{\partial x \partial t} \\ &= \frac{p^2}{a^2}g_1\left(t\right) + \frac{p}{a^2}\left(\frac{\partial u}{\partial t}\right)_0 + \frac{1}{a}g_1'\left(t\right), \end{aligned} \tag{18}$$

and

$$\begin{aligned} \frac{\partial^2 u\left(x_1,t\right)}{\partial x^2} &= \frac{1}{a}\left(p\frac{\partial u\left(x_1,t\right)}{\partial x} + \frac{\partial u\left(x_1,t\right)}{\partial t}\right) = \frac{p}{a}\frac{\partial u\left(x_1,t\right)}{\partial x} + \frac{1}{a}\left(\frac{\partial u}{\partial t}\right)_1 \\ &= \frac{p}{a}g_1\left(t\right) + \frac{ph}{a}\frac{\partial^2 u\left(x_0,t\right)}{\partial x^2} + \frac{ph^2}{2a}\frac{\partial^3 u\left(x_0,t\right)}{\partial x^3} + \frac{1}{a}\left(\frac{\partial u}{\partial t}\right)_1 + O\left(h^3\right). \end{aligned} \tag{19}$$

Substituting Eqs. (17)-(19) into Eq. (1) and rearranging it, the compact difference scheme on the left boundary can be derived from Eq. (11) as:

$$
\begin{aligned}
&\left(\left(\frac{1}{2}-\beta_1\right)\frac{h}{a}+\frac{1}{6}\frac{ph^2}{a^2}+\beta_1\frac{p^2h^3}{2a^3}\right)\left(\frac{\partial u}{\partial t}\right)_0^n+\beta_1\frac{h}{a}\left(\frac{\partial u}{\partial t}\right)_1^n \\
&=-\frac{1}{h}u_0^n+\frac{1}{h}u_1^n+\left(-1-\frac{1}{2}\frac{ph}{a}-\frac{1}{6}\frac{p^2h^2}{a^2}-\beta_1\frac{p^3h^3}{2a^3}\right)g_1(t)+\left(\left(-\frac{1}{6}+\beta_1\right)\frac{h^2}{a}-\beta_1\frac{ph^3}{2a^2}\right)g_1'(t).
\end{aligned}
\tag{20}
$$

The deduction of Eq. (12) for the right boundary follows the same procedure. The expression is given as:

$$
\begin{aligned}
&\beta_2\frac{h}{a}\left(\frac{\partial u}{\partial t}\right)_{N-1}^n+\left(\left(\frac{1}{2}-\beta_2\right)\frac{h}{a}-\frac{1}{6}\frac{ph^2}{a^2}+\beta_2\frac{p^2h^3}{2a^3}\right)\left(\frac{\partial u}{\partial t}\right)_N^n \\
&=\frac{1}{h}u_{N-1}^n-\frac{1}{h}u_N^n+\left(1-\frac{1}{2}\frac{ph}{a}+\frac{1}{6}\frac{p^2h^2}{a^2}-\beta_2\frac{p^3h^3}{2a^3}\right)g_2(t)+\left(\left(\frac{1}{6}-\beta_1\right)\frac{h^2}{a}-\beta_2\frac{ph^3}{2a^2}\right)g_2'(t).
\end{aligned}
\tag{21}
$$

With Eqs. (10), (20) and (21), we have proposed a compact exponential approximation for space grids with a truncation error of $O(h^4)$. Collecting the unknowns at time level $t = n\Delta t$ into a vector $\mathbf{U}(t)$, we get the system of first-order ordinary differential equations given by

$$
\begin{cases}
\mathbf{A}\dfrac{d\mathbf{U}(t)}{dt}=\mathbf{B}\mathbf{U}(t)+\mathbf{g}(t), \\
\mathbf{U}(0)=\varphi_0,
\end{cases}
\tag{22}
$$

where

$$
\mathbf{U}(t)=\left[u_0(t),u_1(t),\ldots,u_{N-1}(t),u_N(t)\right]^T, \tag{23}
$$

$$
\varphi_0=\left[\varphi_0,\varphi_1,\ldots,\varphi_{N-1},\varphi_N\right]^T, \tag{24}
$$

$$
\mathbf{g}(t)=\begin{bmatrix}
\left(-1-\frac{1}{2}\frac{ph}{a}-\frac{1}{6}\frac{p^2h^2}{a^2}-\beta_1\frac{p^3h^3}{2a^3}\right)g_1(t)+\left(\left(-\frac{1}{6}+\beta_1\right)\frac{h^2}{a}-\beta_1\frac{ph^3}{2a^2}\right)g_1'(t) \\
0 \\
\vdots \\
0 \\
\left(1-\frac{1}{2}\frac{ph}{a}+\frac{1}{6}\frac{p^2h^2}{a^2}-\beta_2\frac{p^3h^3}{2a^3}\right)g_2(t)+\left(\left(\frac{1}{6}-\beta_2\right)\frac{h^2}{a}-\beta_2\frac{ph^3}{2a^2}\right)g_2'(t)
\end{bmatrix}. \tag{25}
$$

The matrix $\mathbf{A}$ and $\mathbf{B}$ have an order of $N+1$. The expression of the matrix $\mathbf{A}$ is given as:

$$
\mathbf{A} = \begin{bmatrix}
a_1 & \beta_1 \frac{h}{a} & 0 & \cdots & 0 & 0 \\
\frac{\alpha_2}{h^2} - \frac{\alpha_1}{2h} & 1 - \frac{2\alpha_2}{h^2} & \frac{\alpha_2}{h^2} + \frac{\alpha_1}{2h} & 0 & \cdots & 0 \\
0 & \frac{\alpha_2}{h^2} - \frac{\alpha_1}{2h} & 1 - \frac{2\alpha_2}{h^2} & \frac{\alpha_2}{h^2} + \frac{\alpha_1}{2h} & 0 & \cdots \\
\vdots & \ddots & \ddots & \ddots & & \vdots \\
\cdots & 0 & \frac{\alpha_2}{h^2} - \frac{\alpha_1}{2h} & 1 - \frac{2\alpha_2}{h^2} & 1 - \frac{2\alpha_2}{h^2} & 0 \\
0 & \cdots & 0 & \frac{\alpha_2}{h^2} - \frac{\alpha_1}{2h} & 1 - \frac{2\alpha_2}{h^2} & 1 - \frac{2\alpha_2}{h^2} \\
0 & 0 & \cdots & 0 & \beta_2 \frac{h}{a} & a_{N+1}
\end{bmatrix}, \qquad (26)
$$

where $a_1 = \left(\frac{1}{2} - \beta_1\right)\frac{h}{a} + \frac{1}{6}\frac{ph^2}{a^2} + \beta_1 \frac{p^2 h^3}{2a^3}$ and $a_{N+1} = \left(\frac{1}{2} - \beta_2\right)\frac{h}{a} - \frac{1}{6}\frac{ph^2}{a^2} + \beta_2 \frac{p^2 h^3}{2a^3}$.

The expression of the matrix $\mathbf{B}$ is given as:

$$
\mathbf{B} = \begin{bmatrix}
-\frac{1}{h} & \frac{1}{h} & 0 & \cdots & 0 & 0 \\
\frac{\alpha}{h^2} + \frac{p}{2h} & -\frac{2\alpha}{h^2} & \frac{\alpha}{h^2} - \frac{p}{2h} & 0 & \cdots & 0 \\
0 & \frac{\alpha}{h^2} + \frac{p}{2h} & -\frac{2\alpha}{h^2} & \frac{\alpha}{h^2} - \frac{p}{2h} & 0 & \cdots \\
\vdots & \ddots & \ddots & \ddots & & \vdots \\
\cdots & 0 & \frac{\alpha}{h^2} + \frac{p}{2h} & -\frac{2\alpha}{h^2} & \frac{\alpha}{h^2} + \frac{p}{2h} & 0 \\
0 & \cdots & 0 & \frac{\alpha}{h^2} + \frac{p}{2h} & -\frac{2\alpha}{h^2} & \frac{\alpha}{h^2} - \frac{p}{2h} \\
0 & 0 & \cdots & 0 & \frac{1}{h} & -\frac{1}{h}
\end{bmatrix}. \qquad (27)
$$

### 2.3 Time discretization

From Lemma 2 in Appendix B, we have the following results:

$$
\begin{cases}
|a_1| > \left|\beta_1 \dfrac{h}{a}\right|, \\
|a_{N+1}| > \left|\beta_2 \dfrac{h}{a}\right|, \\
\left|1 - \dfrac{2\alpha_2}{h^2}\right| > \left|\dfrac{\alpha_2}{h^2} - \dfrac{\alpha_1}{2h}\right| + \left|\dfrac{\alpha_2}{h^2} + \dfrac{\alpha_1}{2h}\right|.
\end{cases} \tag{28}
$$

This proves that matrix **A** is strictly diagonally dominant. From Lemma 2, matrix A is nonsingular. Therefore, Eq. (22) can be rewritten as:

$$
\begin{cases}
\dfrac{d\mathbf{U}(t)}{dt} = \mathbf{A}^{-1}\mathbf{B}\mathbf{U}(t) + \mathbf{A}^{-1}\mathbf{g}(t), \\
\mathbf{U}(0) = \varphi_0,
\end{cases} \tag{29}
$$

Note that the Taylor series of $\mathbf{U}(t)$ at time $T = t_n + \Delta t$ can be expressed as:

$$
\mathbf{U}(t_n + \Delta t) = \left(1 + \Delta t \frac{d}{dt} + \frac{\Delta t^2}{2!}\frac{d^2}{dt^2} + \ldots\right)\mathbf{U}(t) = \exp\left(\Delta t \frac{d}{dt}\right)\mathbf{U}(t). \tag{30}
$$

Using the Padé approximation for time discretization:

$$
\exp\left(\Delta t \frac{d}{dt}\right) = \left[1 - \frac{1}{2}\Delta t \frac{d}{dt} + \frac{1}{12}\left(\Delta t \frac{d}{dt}\right)^2\right]^{-1}\left[1 + \frac{1}{2}\Delta t \frac{d}{dt} + \frac{1}{12}\left(\Delta t \frac{d}{dt}\right)^2\right], \tag{31}
$$

Eq. (30) can be rewritten as:

$$
\mathbf{U}(t_n + \Delta t) = \mathbf{L}^{-1}\left(\mathbf{H}\mathbf{U}(t_n) + \mathbf{L}_1\mathbf{A}^{-1}g(t_{n+1}) + \mathbf{H}_1\mathbf{A}^{-1}g(t_n)\right) + \frac{1}{12}\Delta t^2\mathbf{L}^{-1}\mathbf{A}^{-1}\left[g'(t_{n+1}) - g'(t_n)\right], \tag{32}
$$

Where,

$$
\begin{aligned}
\mathbf{L} &= \mathbf{I} - \frac{1}{2}\Delta t\mathbf{A}^{-1}\mathbf{B} + \frac{1}{12}\left(\Delta t\mathbf{A}^{-1}\mathbf{B}\right)^2, \mathbf{L}_1 = \frac{1}{2}\Delta t\mathbf{A}^{-1}\mathbf{B} - \frac{1}{12}\left(\Delta t\mathbf{A}^{-1}\mathbf{B}\right)^2, \\
\mathbf{H} &= \mathbf{I} + \frac{1}{2}\Delta t\mathbf{A}^{-1}\mathbf{B} + \frac{1}{12}\left(\Delta t\mathbf{A}^{-1}\mathbf{B}\right)^2, \mathbf{H}_1 = \frac{1}{2}\Delta t\mathbf{A}^{-1}\mathbf{B} + \frac{1}{12}\left(\Delta t\mathbf{A}^{-1}\mathbf{B}\right)^2.
\end{aligned} \tag{33}
$$

This is the high-order exponential scheme for solving the 1D unsteady convection-diffusion equation with Neumann boundary conditions. The analysis shows that it achieves a fourth-order accuracy in all spatial grid points, and it has fourth-order accuracy in temporal grids with the Padé approximation.

## 3 Stability analysis

For the finite difference scheme, the amplification matrix is given by:

$$
\mathbf{\Phi} = \left[\mathbf{I} - \frac{1}{2}\Delta t\mathbf{A}^{-1}\mathbf{B} + \frac{1}{12}\left(\Delta t\mathbf{A}^{-1}\mathbf{B}\right)^2\right]^{-1}\left[\mathbf{I} + \frac{1}{2}\Delta t\mathbf{A}^{-1}\mathbf{B} + \frac{1}{12}\left(\Delta t\mathbf{A}^{-1}\mathbf{B}\right)^2\right]. \tag{34}
$$

Let $\lambda_i$ be the eigenvalues of the matrix $\mathbf{A}^{-1}\mathbf{B}$, the eigenvalues of matrix $\mathbf{\Phi}$ can be calculated as:

$$
\lambda(\mathbf{\Phi}) = \left[1 - \frac{1}{2}\Delta t\lambda_i + \frac{1}{12}(\Delta t\lambda_i)^2\right]^{-1}\left[1 + \frac{1}{2}\Delta t\lambda_i + \frac{1}{12}(\Delta t\lambda_i)^2\right]. \tag{35}
$$

If the real part of the eigenvalue $\mathrm{Re}(\lambda_i)$ is negative, the scheme can be proven to be unconditional stable[11]:

$$\max_{\lambda_i}\left|\frac{1+\frac{1}{2}\Delta t\lambda_i+\frac{1}{12}(\Delta t\lambda_i)^2}{1-\frac{1}{2}\Delta t\lambda_i+\frac{1}{12}(\Delta t\lambda_i)^2}\right|_i \le 1. \tag{36}$$

Due to the complexity of the Neumann boundary conditions, the distribution of the eigenvalue $\lambda_i$ of the matrix $\mathbf{A}^{-1}\mathbf{B}$ is investigated numerically by varying step size *h* and the *p/a* ratio. The product of *h* and *p/a* is defined as Péclet Number (*Pe*):

$$Pe=\frac{ph}{a}. \tag{37}$$

which is defined as the ratio of mass transport contributed by convection to those by diffusion. Dividing Eq.(1) by *a*, one has

$$\frac{1}{a}\frac{\partial u}{\partial t}+\frac{p}{a}\frac{\partial u}{\partial x}=\frac{\partial^2 u}{\partial x^2}. \tag{38}$$

Let $p'=p/a$ and $a'=1$, the Eq. (38) will have the form of

$$\frac{1}{a}\frac{\partial u}{\partial t}+p'\frac{\partial u}{\partial x}=a'\frac{\partial^2 u}{\partial x^2}. \tag{39}$$

Following the same procedure for spatial discretization, the PDE system can be approximated as

$$\begin{cases}\dfrac{d\mathbf{U}(t)}{dt}=a\mathbf{A}'^{-1}\mathbf{B}'\mathbf{U}(t)+a\mathbf{A}'^{-1}\mathbf{g}(t),\\ \mathbf{U}(0)=\varphi_0,\end{cases} \tag{40}$$

where $\mathbf{A}'$ and $\mathbf{B}'$ share the same expressions as $\mathbf{A}$ and $\mathbf{B}$ ,except that $p$ is replaced by $p'$ and $a$ by $a'$. By comparing Eq. (29) and Eq. (40), it can be found that the eigenvalues of $\mathbf{A}'^{-1}\mathbf{B}'$ have the same sign of matrix $\mathbf{A}^{-1}\mathbf{B}$ with a scaling factor of *a*. Since $a'$ is a constant, the matrix $\mathbf{A}'^{-1}\mathbf{B}'$ contains only two variables. One is the step size *h* and the another is the Péclet Number or $p'$. In order to evaluate the eigenvalue distribution of $\mathbf{A}'^{-1}\mathbf{B}'$, a negative eigenvalue weight *R* is introduced to quantify the absolute ratio of negative eigenvalues among all the eigenvalues given by $\mathbf{A}'^{-1}\mathbf{B}'$. The definition of *R* is given as

$$R=\frac{\sum_{i,\lambda_i<0}\left|\mathrm{Re}(\lambda_i)\right|}{\sum_i\left|\mathrm{Re}(\lambda_i)\right|}. \tag{41}$$

The distribution of *R* is plotted in log scale for visualization in Fig. 1. To generate the matrix $\mathbf{A}'^{-1}\mathbf{B}'$, the computational spatial domain is given as *L* = 1. The step size *h* ranges from 0.001 to 1 and the *p/a* ratio ranges from 0.001 to $10^4$. As can be seen in the figure, all the real part of eigenvalues are negative for convection dominated problems with $Pe\ge 1$, which yields *R* = 1 in this case. This indicates that the developed scheme is unconditionally stable for convection dominated problems. For smaller *Pe*, it can be seen that most of the region retains the unconditionally stable property. At the top left corner, with *Pe* around 0.0001, positive real eigenvalues exist in certain areas. In this region, the scheme should be used with caution, as positive real eigenvalues can be the root of instability.

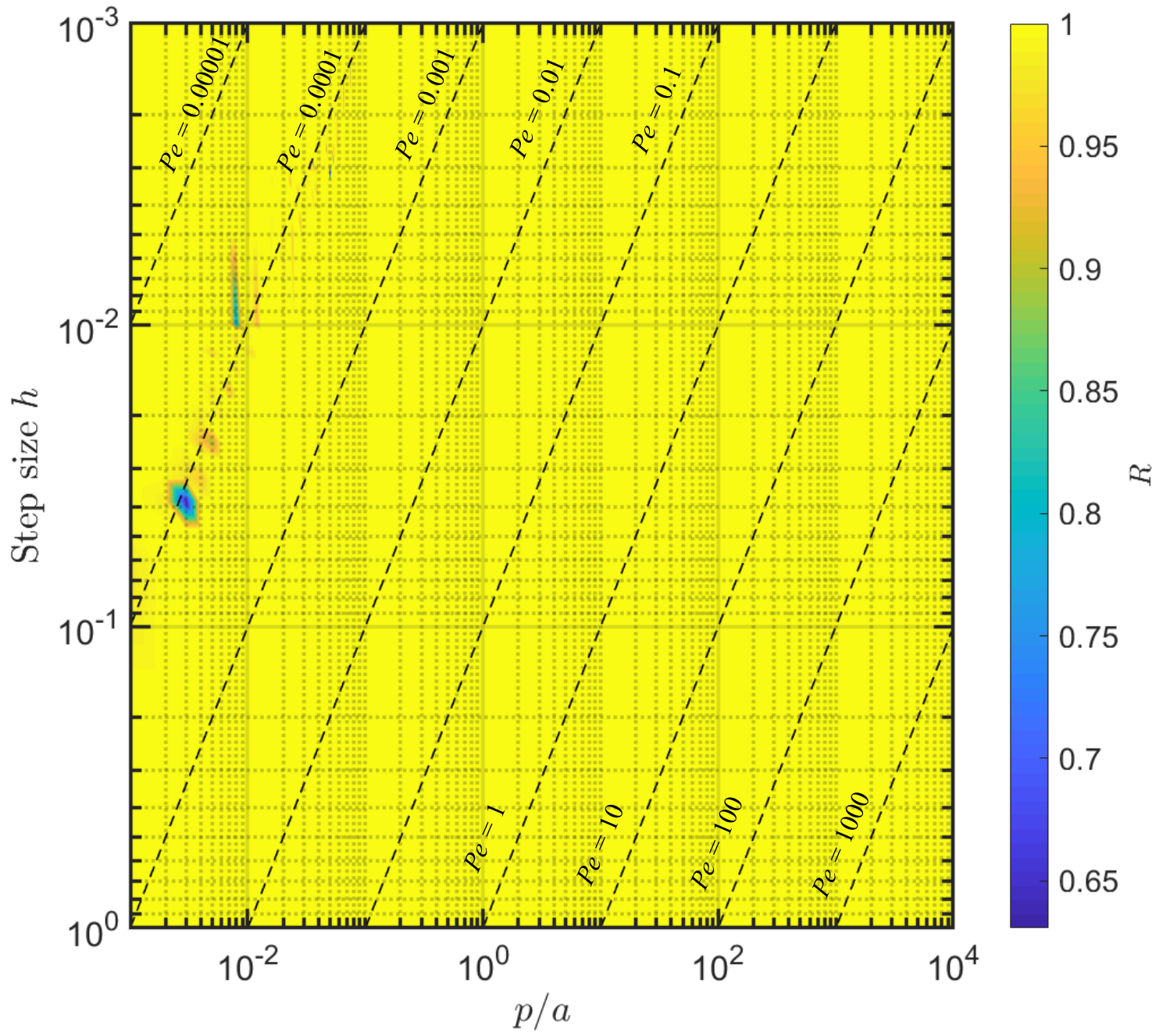

Fig. 1 The distribution of absolute negative eigenvalue ratio $R$ of the matrix $\mathbf{A}'^{-1}\mathbf{B}'$ with respect to varying spatial step size $h$ and $p/a$ ratio. The dash lines indicate the value of Péclet Number.

## 4 Numerical experiments

In this section, four partial differential equations with known exact solutions are used to test the accuracy and the effectiveness of the newly proposed method for solving the 1D unsteady convection-diffusion equation with the Neumann boundary condition. The implicit Crank-Nicolson method $O\left(\Delta t^2 + h^2\right)$ and the Cao's[18] method $O\left(\Delta t^4 + h^4\right)$ are compared with the present scheme $O\left(\Delta t^4 + h^4\right)$. In the following problems, the Courant number is defined as $\lambda = \frac{p\Delta t}{h}$, and the diffusion parameter $s = \frac{a\Delta t}{h^2}$ is used. The Péclet Number is defined previously in Eq. (37).

### 4.1 Problem 1

Consider the following convection-diffusion equation [22]:

$$\frac{\partial u}{\partial t} + p\frac{\partial u}{\partial x} = a\frac{\partial^2 u}{\partial x^2}, 0 \le x \le 2. \tag{42}$$

with boundary conditions:

$$\frac{\partial u(0,t)}{\partial x} = \frac{p}{2a\sqrt{1+t}} \exp\left\{ -\frac{(1+t)^2 p^2}{4a(1+t)} \right\}, \text{ and}$$

$$\frac{\partial u(2,t)}{\partial x} = \frac{-2+(1+t)p}{2a(1+t)\sqrt{1+t}} \exp\left\{ -\frac{(x-(1+t)p)^2}{4a(1+t)} \right\}. \tag{43}$$

The initial condition of the problem is a Gaussian pulse centered at $x=p$. The analytical solution to this equation is given as:

$$u(x,t) = \frac{1}{\sqrt{1+t}} \exp\left\{ -\frac{(x-(1+t)p)^2}{4a(1+t)} \right\}, t > 0\,. \tag{44}$$

To test the accuracy of the present scheme, the convergence rate is first presented with changing $h$ and $\Delta t$ using $p$ = 0.25, $a$ = 0.1. In Table 1, different values of $h$ are selected to test the fourth-order accuracy of the spatial scheme with fixed time step $\Delta t$ = 0.01. The $L^2$-norm error and the convergence rate are given at $T$ = 2. The convergence rate is calculated by $\log_2(E_{L1}/E_{L2})$, where $E_{L1}$ and $E_{L2}$ are an error with the grid size of $h$ and $h/2$, respectively. The convergence rates are approximately 4 for the present scheme and the Cao scheme. In Table 2, the time accuracy is checked by decreasing $\Delta t$ with fixed $h$ = 0.01 at $T$ = 10. The convergence rate of time is calculated by $\log_2(E_{T1}/E_{T2})$, where $E_{T1}$ and $E_{T2}$ are errors with the grid sizes of $\Delta t$ and $\Delta t/2$, respectively. Table 3 confirms that the present method has an overall fourth-order convergence rate by reducing $h$ and $\Delta t$ at the same time. These three tables show that Cao's method and the present scheme achieve a fourth-order convergence rate in both time and space as expected. The Crank-Nicolson scheme did not achieve second-order convergence rate for this problem.

To test the suitability of the present scheme for convection dominated problems, Fig. 2 show the solutions of two different Péclet numbers with $h$ = 0.01 and $\Delta t$ = 0.01 at $T$ = 2. For a small $Pe$ of 0.025, the Crank-Nicolson method does not yield an exact solution. For a large $Pe$ number at 25, both the Crank-Nicolson method and Cao's scheme show an oscillation at the bottom of the Gaussian pulse. The present method fits the analytical solution very well for both the small $Pe$ and for the convection dominated condition with large $Pe$ number.

Table 1 Spatial $L^2$-norm error and convergence rate for solving Problem 1 with $\Delta t$ =0.01 at $T$=2.

| $h$ | Crank-Nicolson | | Cao | | Present | |
|---|---|---|---|---|---|---|
| | $L^2$-norm | Rate | $L^2$-norm | Rate | $L^2$-norm | Rate |
| 0.040 | 8.74(-2) | | 3.09(-7) | | 2.08(-7) | |
| 0.020 | 4.25(-2) | - | 1.93(-8) | 4.000 | 1.30(-8) | 3.998 |
| 0.010 | 6.22(-2) | - | 1.21(-9) | 3.993 | 8.09(-10) | 4.005 |
| 0.005 | 1.24(-1) | - | 8.74(-11) | 3.793 | 4.03(-11) | 4.327 |

Table 2 Temporal $L^2$-error and convergence rate for solving Problem 1 with $h$ = 0.01 at $T$ = 2.

| $\Delta t$ | Crank-Nicolson | | Cao | | Present | |
|---|---|---|---|---|---|---|
| | $L^2$-norm | Rate | $L^2$-norm | Rate | $L^2$-norm | Rate |
| 1.000 | 3.21(-1) | | 2.10(-3) | | 6.66(-4) | |
| 0.500 | 2.98(-1) | - | 2.07(-4) | 3.344 | 4.77(-5) | 3.801 |
| 0.250 | 2.77(-1) | - | 1.40(-5) | 3.887 | 3.13(-6) | 3.930 |
| 0.125 | 2.49(-1) | - | 8.42(-7) | 4.052 | 1.98(-7) | 3.981 |

Table 3 $L^2$-error and convergence rate for solving Problem 1 with both changing $\Delta t$ and $h$.

| $\Delta t/h$ | Crank-Nicolson | | Cao | | Present | |
|---|---|---|---|---|---|---|
| | $L^2$-norm | Rate | $L^2$-norm | Rate | $L^2$-norm | Rate |
| 1.000/0.040 | 2.78(-1) | | 2.10(-3) | | 6.78(-4) | |
| 0.500/0.020 | 2.74(-1) | - | 2.07(-4) | 3.340 | 4.80(-5) | 3.821 |
| 0.250/0.010 | 2.77(-1) | - | 1.39(-5) | 3.892 | 3.13(-6) | 3.937 |
| 0.125/0.005 | 1.37(-1) | - | 8.40(-7) | 4.055 | 1.99(-7) | 3.979 |

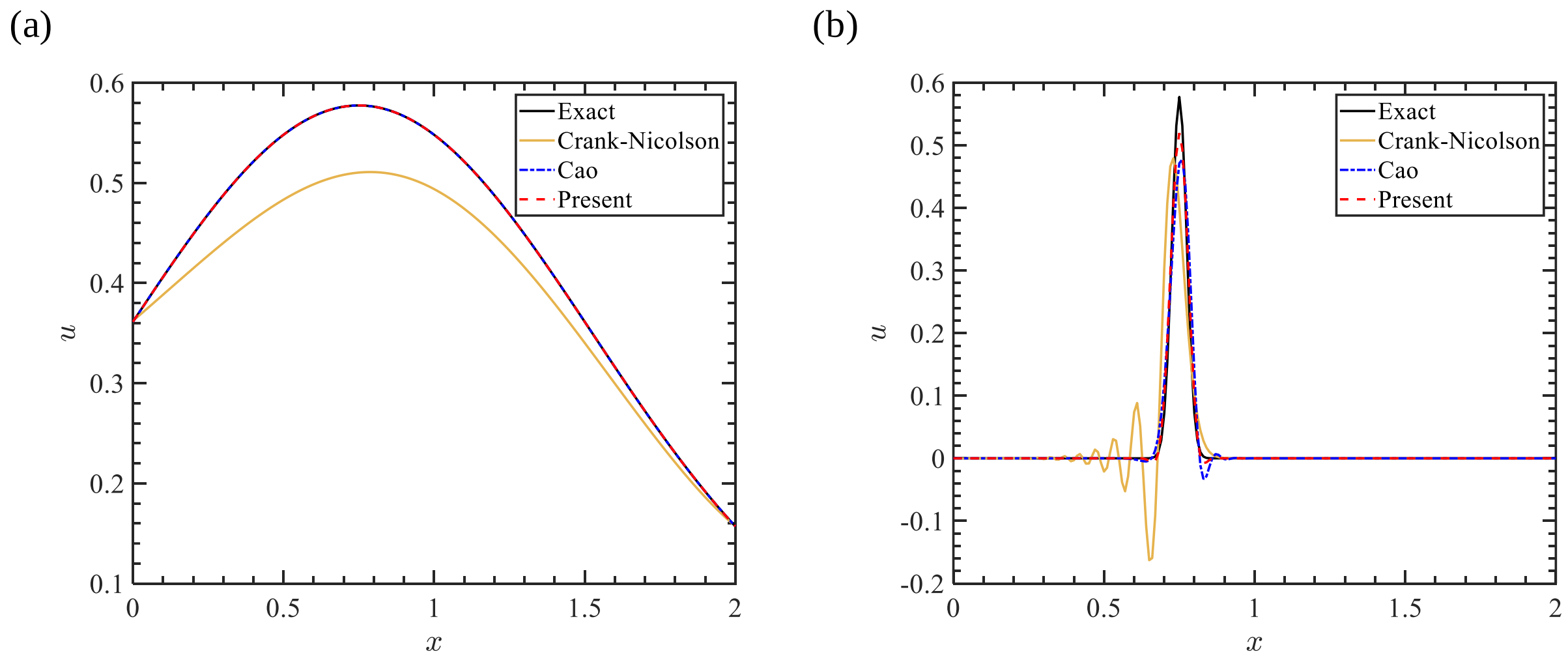

Fig. 2 Comparison of different numerical schemes for solving Problem 1 with $h$ = 0.01 and $\Delta t$ =0.01 at $T$ = 2 using (a) $Pe$ =0.025 and (b) $Pe$ =25.

## 4.2 Problem 2

In this case, the following convection-diffusion equation for a unit square wave is considered:

$$\begin{cases} \dfrac{\partial u}{\partial t} + p\dfrac{\partial u}{\partial x} = a\dfrac{\partial^2 u}{\partial x^2}, 0 \le x \le 1, t > 0, \\ u(x,0) = \begin{cases} 1, & 0.0905 \le x \le 0.205, \\ 0, & \text{otherwise}, \end{cases} \\ \dfrac{\partial u(0,t)}{\partial x} = 0, \dfrac{\partial u(1,t)}{\partial x} = 0, t > 0. \end{cases} \tag{45}$$

The analytical solution of this problem is given as [13]:

$$u(x,t)=\frac{1}{2}\left[erf\left(\frac{0.205-x+pt}{2\sqrt{at}}\right)+erf\left(\frac{-0.0905+x-pt}{2\sqrt{at}}\right)\right], \tag{46}$$

where the error function is defined as $erf(x)=\frac{2}{\sqrt{\pi}}\int_0^x e^{-\eta^2}d\eta$.

In Table 4, the $L^2$-norm and $L^\infty$-norm errors are listed using $Pe$ = 100 with $h$ = 0.01 and $\Delta t$ = 0.0001 at $T$ = 0.006. Four Courant numbers $\lambda$ = 0.2, 0.4, 0.6, 0.8 are used in the test. The $L^2$-norm indicates the overall accuracy of the scheme over the entire spatial domain. The $L^\infty$-norm represents the maximum difference between the numerical result and the analytical solution. In this example, the maximum of the difference is usually found at the front or rear edge of the square wave. The results in Table 4 shows that the newly proposed method has an overall good accuracy in capturing the square wave movement in the spatial domain. The $L^2$-norm and $L^\infty$-norm errors are smaller than the Crank-Nicolson method and the Cao's method with all four Courant numbers.

Fig. 3 visualizes the numerical solutions of different methods for $Pe$ = 10, 100, 1000, 10000 with $h$ = 0.01 and $\Delta t$ =0.0001 at time $T$ = 0.006. The Courant number is chosen as 0.5. As can be seen in the figure, oscillation and numerical diffusion of all the schemes will increase due to the increase of $Pe$. The Crank-Nicolson method shows an obvious oscillation on the left hand side of the square wave, whereas the Cao's method shows a more significant oscillation on the right hand side. The present method shows a much smaller magnitude of oscillation and over diffusion compared to those two schemes, and it is less affected by the increasing of $Pe$ number. This indicates that the present scheme can be a good option for calculating the convection dominated problems.

Table 4 Comparison of $L^2$-norm and $L^\infty$-norm error for solving Problem 2 at time $T$ = 0.006 with $h$ = 0.01 and $\Delta t$ =0.0001 using $Pe$ = 100.

| λ | Crank-Nicolson | | Cao | | Present | |
|---|---|---|---|---|---|---|
| | $L^2$-norm | $L^\infty$-norm | $L^2$-norm | $L^\infty$-norm | $L^2$-norm | $L^\infty$-norm |
| 0.2 | 0.0943 | 0.4037 | 0.0830 | 0.4705 | 0.0687 | 0.3310 |
| 0.4 | 0.1176 | 0.4547 | 0.0838 | 0.4350 | 0.0687 | 0.2977 |
| 0.6 | 0.1171 | 0.3950 | 0.0872 | 0.3980 | 0.0644 | 0.2646 |
| 0.8 | 0.1100 | 0.4046 | 0.0874 | 0.3683 | 0.0690 | 0.2386 |

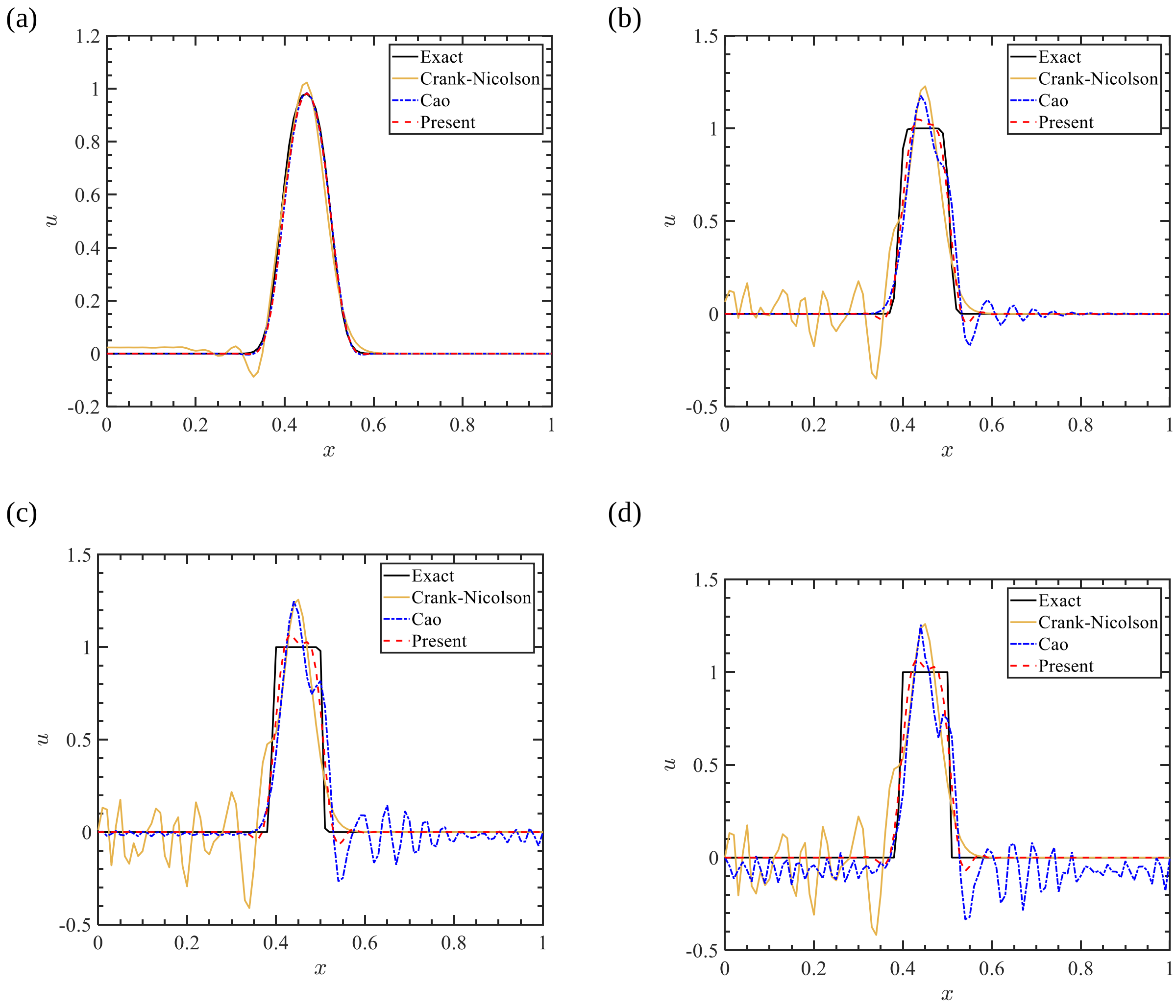

Fig. 3 Comparison of different numerical schemes for solving Problem 2 with $h$ = 0.01, $\Delta t = 0.001$, at $T$ = 0.006 using (a) $Pe$ = 10, (b) $Pe$ = 100, (c) $Pe$ = 1000, (d) $Pe$ = 10000.

## 4.3 Problem 3

Consider the following convection-diffusion equation:

$$\begin{cases} \dfrac{\partial u}{\partial t}+\dfrac{\partial u}{\partial x}=\dfrac{1}{\mathrm{Re}}\dfrac{\partial^2 u}{\partial x^2}, 0\le x\le 1, t>0, \\ u(x,0)=0, 0<x\le 1,\ u(x,0)=1, x=0, \\ \dfrac{\partial u(0,t)}{\partial x}=0, \dfrac{\partial u(0,t)}{\partial x}=0, t>0, \end{cases} \tag{47}$$

the analytical solution to the problem is given as:

$$u(x,t)=\frac{1}{2}\left[erfc\left(\frac{x-pt}{2\sqrt{at}}\right)+e^{\frac{px}{a}}erfc\left(\frac{x+pt}{2\sqrt{at}}\right)\right], \tag{48}$$

where the complementary error function $erfc(x)$ is defined as:

$$erfc\left(x\right)=\frac{2}{\sqrt{\pi}}\int_{x}^{\infty}e^{-t^{2}}dt. \tag{49}$$

This equation simulates the propagation of a one-dimensional shock wave, which has a sudden change at the wave front. The numerical solutions of the present scheme, the Cao's method and the Crank-Nicolson method are compared with $\Delta t$ =0.0001 and $h$ = 0.01 at $T$ = 0.006. Table 5 gives the $L^2$-norm and $L^\infty$-norm error where $Pe$ = 100. Four different Courant numbers are used for comparison in the table. Fig. 4 shows the propagation of the shock wave in time sequence from $t$ = 0 to $t$ = 0.8 with $p$ = 1, $a$ = 0.0001. The time grid size is $dt$ = 0.0001 and the space grid size is $h$ = 0.01. Compared to the analytical solution shown in Fig. 4 (a), the present method shows a well behaved result in simulating the shock wave propagation. No obvious oscillation is observed. For Cao's method, there are ripples shown in Fig. 4 (c) with the propagation of the shock wave. The Crank-Nicolson method shows an unphysical oscillation in the upstream of shockwave as shown in Fig. 4 (b).

Table 5 Comparison of $L^2$-norm and $L^\infty$-norm error for solving Problem 3 at time $T$ = 0.006 with $h$ = 0.01 and $\Delta t$ =0.0001 using $Pe$ = 100.

| λ | Crank-Nicolson | | Cao | | Present | |
|---|---|---|---|---|---|---|
| | $L^2$-norm | $L^\infty$-norm | $L^2$-norm | $L^\infty$-norm | $L^2$-norm | $L^\infty$-norm |
| 0.2 | 0.0552 | 0.3835 | 0.0643 | 0.4766 | 0.0389 | 0.3267 |
| 0.4 | 0.0661 | 0.3919 | 0.0651 | 0.4466 | 0.0369 | 0.2966 |
| 0.6 | 0.0734 | 0.3786 | 0.0648 | 0.4115 | 0.0349 | 0.2643 |
| 1.0 | 0.0869 | 0.3737 | 0.0640 | 0.3587 | 0.0319 | 0.2166 |

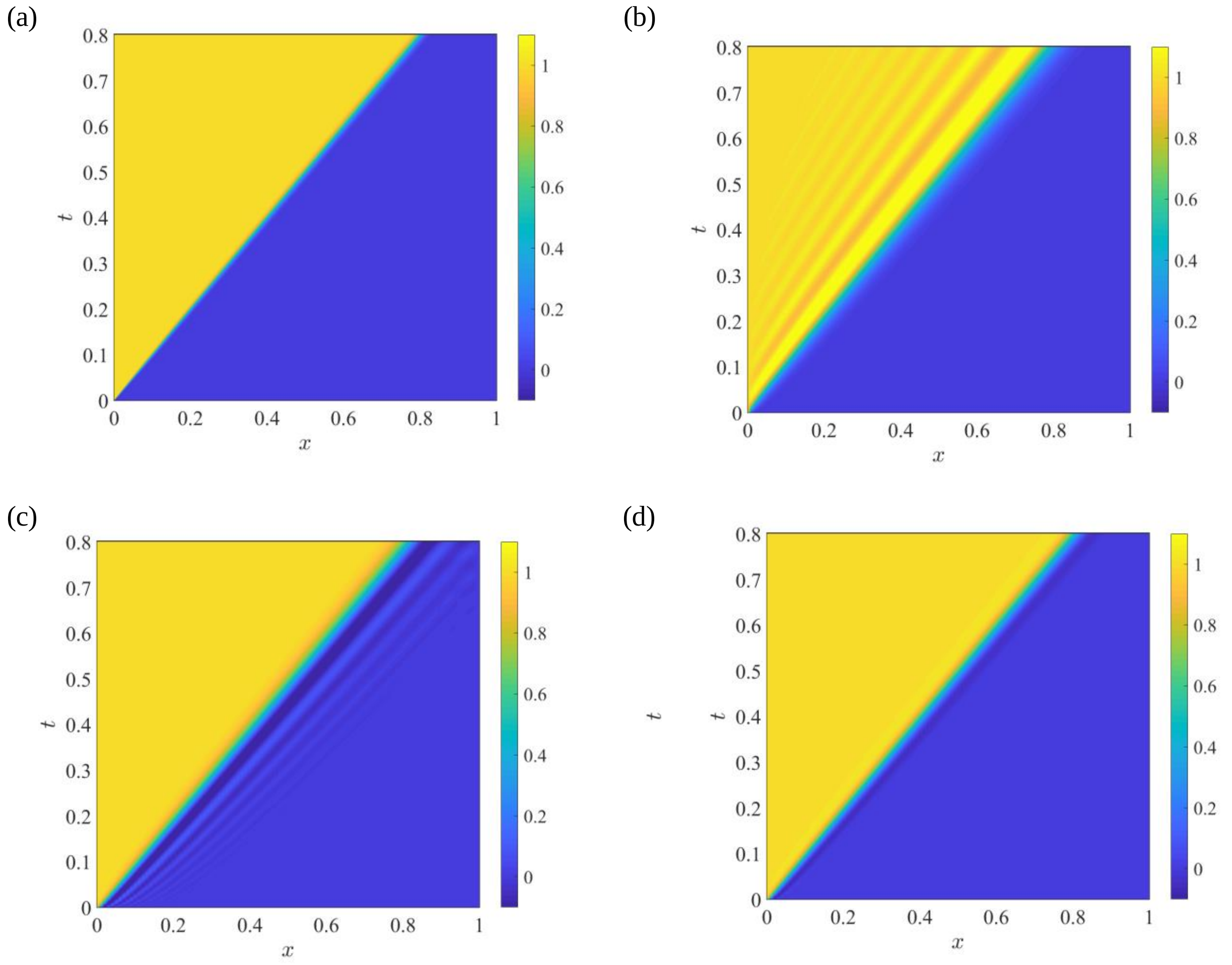

Fig. 4 Comparison of (a) analytical solution, (b) Crank-Nicolson method, (c) Cao's method and (d) the present method for simulating the shock wave propagation with $p$ = 1, $a$ = 0.0001, $dt$ = 0.0001, $h$ = 0.01 from $T$ = 0 to 0.8.

### 4.4 Problem 4

Consider the following equation [23]:

$$\begin{cases} \dfrac{\partial u}{\partial t}+\dfrac{\partial u}{\partial x}=\dfrac{1}{\mathrm{Re}}\dfrac{\partial^2 u}{\partial x^2}, 0\le x\le 1, t>0, \\ u(x,0)=0, 0<x<1, \\ \dfrac{\partial u(0,t)}{\partial x}=\dfrac{\mathrm{Re}}{e^{\mathrm{Re}}-1}+\displaystyle\sum_{m=1}^{\infty}\dfrac{(-1)^m (m\pi)^2}{2(m\pi)^2+\dfrac{\mathrm{Re}^2}{2}}e^{-\frac{\mathrm{Re}}{2}-\left[\frac{(m\pi)^2}{\mathrm{Re}}+\frac{\mathrm{Re}}{4}\right]t}, u(1,t)=1, t>0, \end{cases} \tag{50}$$

the analytical solution can be obtained by using the separation of variables method as:

$$u(x,t)=\frac{e^{\mathrm{Re}\,x}-1}{e^{\mathrm{Re}}-1}+\sum_{m=1}^{\infty}\frac{(-1)^m m\pi}{2(m\pi)^2+\frac{\mathrm{Re}^2}{2}}e^{\frac{\mathrm{Re}(x-1)}{2}}\sin(m\pi x)e^{-\left[\frac{(m\pi)^2}{\mathrm{Re}}+\frac{\mathrm{Re}}{4}\right]t}. \tag{51}$$

This problem has a boundary layer on the right side with the increasing *Re*, which causes an initial discontinuity at $x = 1$. Thus, mixed boundary conditions are present for this problem.

To check the influence of spatial resolution on the Crank-Nicolson method, the Cao's method and the present method, different step sizes $h = 0.04$ and $h = 0.1$ are used for test under Reynolds number of $Re = 100$ and 10000. The results are shown in Fig. 5. The Cao's method shows an over diffusion and cannot capture the boundary layer at $x = 1$. The Crank-Nicolson scheme shows an strong oscillation and it is suppressed by decreasing the space step size and *Re*. Table 6 shows the $L^2$-norm and $L^\infty$-norm errors with a coarse mesh ($h = 0.1$) for different Reynolds numbers. With the coarse mesh setup, the Crank-Nicolson method and Cao's method show large errors with the increasing of *Re* due to their relative low resolution in spatial domain. The present method has a higher spatial resolution and can yield a satisfactory result even for high *Re* conditions.

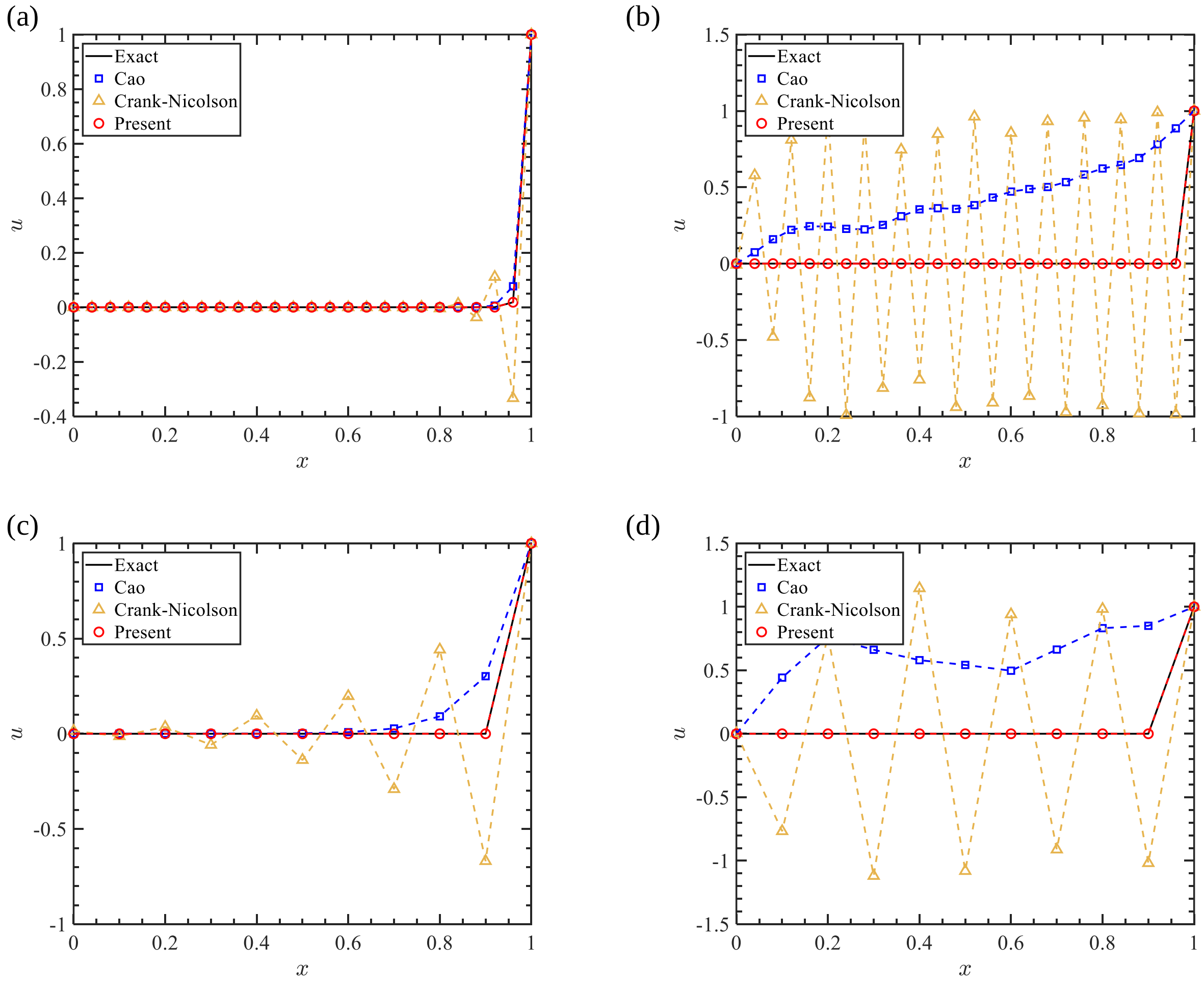

Fig. 5 Comparison of different numerical schemes for solving Problem 4 at time $T = 1$ with $\Delta t = 0.01$ using (a) $h = 0.04$, $Re = 100$, (b) $h = 0.04$, $Re = 10000$, (c) $h = 0.1$, $Re = 100$ and (d) $h = 0.1$, $Re = 10000$.

Table 6 Comparison of $L^2$-norm and $L^\infty$-norm error for solving Problem 4 at time $T = 1$ with $h = 0.1$ and $\Delta t = 0.01$.

| *Re* | Crank-Nicolson | | Cao | | Present | |
|---|---|---|---|---|---|---|
| | $L^2$-norm | $L^\infty$-norm | $L^2$-norm | $L^\infty$-norm | $L^2$-norm | $L^\infty$-norm |
| 10 | 1.40(-2) | 3.44(-2) | 2.34(-4) | 5.53(-4) | 1.12(-4) | 3.45(-6) |
| 100 | 2.83(-1) | 6.69(-1) | 1.00(-1) | 3.02(-1) | 1.27(-6) | 4.24(-11) |
| 1000 | 7.85(-1) | 9.80(-1) | 4.13(-1) | 7.93(-1) | 0 | 0 |
| 10000 | 9.30(-1) | 1.15(0) | 6.28(-1) | 8.50(-1) | 0 | 0 |
| 100000 | 9.46(-1) | 1.17(0) | 7.07(-1) | 1.03(0) | 0 | 0 |

## 5 Conclusion

In this paper, a high-order exponential scheme has been proposed for solving 1D unsteady convection-diffusion equations with Neumann boundary condition. The proposed method derives a compact exponential scheme for spatial discretization in both interior and boundary grids. This numerical method is proved to be unconditionally stable for the convection dominated problem. Through the numerical examples, this scheme shows a robust behavior for the convection dominated problems compared to the Crank-Nicolson method and the Cao's method. In our future works, we will extend this scheme for 2D and 3D unsteady convection-diffusion problems with Neumann boundary conditions.

**Appendix A.** The interior spatial points discretization

For the 1D steady-state convection-diffusion equation:

$$-a\frac{\partial^2 u}{\partial x^2} + p\frac{\partial u}{\partial x} = f, \tag{A.1}$$

it can be rearranged as:

$$-a\frac{\partial}{\partial x}\left(e^{-\frac{px}{a}}\frac{\partial u}{\partial x}\right) = e^{-\frac{px}{a}} f. \tag{A.2}$$

Integrate Eq. (A.2) over the domain $[x_{i-\frac{1}{2}}, x_{i+\frac{1}{2}}]$ gives:

$$-a\left[\left(e^{-\frac{px}{2a}}\frac{\partial u}{\partial x}\right)_{i+\frac{1}{2}} - \left(e^{-\frac{px}{2a}}\frac{\partial u}{\partial x}\right)_{i-\frac{1}{2}}\right] = \int_{x_{i-\frac{1}{2}}}^{x_{i+\frac{1}{2}}} e^{-\frac{px}{a}} f dx. \tag{A.3}$$

To replace the derivatives on the left side of Eq. (A.3), the $\frac{\partial u_i}{\partial x}$ term is approximated using the central difference scheme:

$$\frac{\partial u_i}{\partial x} = \frac{u_{i+\frac{1}{2}} - u_{i-\frac{1}{2}}}{h} + O\left(h^2\right). \tag{A.4}$$

Thus, we have:

$$-\alpha\delta_x^2 u_i + p\delta_x u_i = \frac{e^{\frac{px_i}{a}}}{e^{\frac{ph}{2a}} - e^{-\frac{ph}{2a}}} \int_{x_{i-\frac{1}{2}}}^{x_{i+\frac{1}{2}}} e^{-\frac{px}{a}} f dx, \tag{A.5}$$

where,

$$\alpha = \begin{cases} \frac{ph}{2}\coth\left(\frac{ph}{2a}\right), p \neq 0, \\ a, \qquad\qquad p = 0. \end{cases} \tag{A.6}$$

Using a Taylor series expansion, the source term $f(x)$ can be expressed as:

$$f = f_i + (x - x_i)\delta_x f_i + \frac{(x-x_i)^2}{2!}\delta_x^2 f_i + \frac{(x-x_i)^3}{3!}\delta_x^3 f_i(\xi_i) + O\left((x-x_i)^4\right),\ \xi_i \in (x_{i-1}, x_{i+1}) \tag{A.7}$$

where,

$$\delta_x f_i = \frac{f(x_{i+1}) - f(x_{i-1})}{2h}, \delta_x^2 f_i = \frac{f(x_{i+1}) - 2f(x_i) + f(x_{i-1})}{h^2}, \tag{A.8}$$

for $x \in (x_{i-1}, x_{i+1}), i = 1, 2, ..., N-1$. Replace right-hand side Eq.(A.5) by Eq.(A.7) and rearrange it, we have:

$$-\alpha\delta_x^2 u_i + p\delta_x u_i = f_i + \alpha_1\delta_x f_i + \alpha_2\delta_x^2 f_i + O\left(h^4\right), \tag{A.9}$$

where the first order and second order central difference operators $\delta_x u_i$ and $\delta_x^2 u_i$ are defined, respectively, as:

$$\delta_x u_i = \frac{u_{i+1} - u_{i-1}}{2h}, \text{ and } \delta_x^2 u_i = \frac{u_{i+1} - 2u_i + u_{i-1}}{h^2}. \tag{A.10}$$

Omitting the truncation error, we obtain the following three-point fourth-order compact finite difference formulation for Eq.(6):

$$-\alpha\delta_x^2 u_i + p\delta_x u_i = f_i + \alpha_1\delta_x f_i + \alpha_2\delta_x^2 f_i, \tag{A.11}$$

where,

$$\alpha = \begin{cases} \frac{ph}{2}\coth\left(\frac{ph}{2a}\right), p \neq 0, \\ a, \qquad\qquad p = 0, \end{cases} \quad \alpha_1 = \begin{cases} \frac{a-\alpha}{p}, p \neq 0, \\ 0, \qquad p = 0, \end{cases} \quad \alpha_2 = \begin{cases} \frac{a(a-\alpha)}{p^2} + \frac{h^2}{6}, p \neq 0, \\ \frac{h^2}{12}, \qquad p = 0. \end{cases} \tag{A.12}$$

A detailed analysis of this compact scheme for the interior spatial points can be found in the previous work[13]. The scheme is provided in the appendix for the completeness of the spatial scheme derivation.

## Appendix B. Proof the strictly diagonal dominance of matrix A

**Lemma 1.** [24] If the matrix $\boldsymbol{\Psi}$ is a real, strictly diagonally dominant, then $\boldsymbol{\Psi}$ is nonsingular.

**Lemma 2.** [13] Assume that $\alpha$, $\alpha_1$ and $\alpha_2$ are given by Eq. (9), then the following relations will hold:

$$1 - \frac{2\alpha_2}{h^2} > 0; -\frac{2\alpha}{h^2} < 0, \tag{B.1}$$

$$\left|1-\frac{2\alpha_2}{h^2}\right| > \left|\frac{\alpha_2}{h^2}-\frac{\alpha_1}{2h}\right| + \left|\frac{\alpha_2}{h^2}+\frac{\alpha_1}{2h}\right|, \tag{B.2}$$

$$\left|-\frac{2\alpha}{h^2}\right| = \left|\frac{\alpha}{h^2}+\frac{p}{2h}\right| + \left|\frac{\alpha}{h^2}-\frac{p}{2h}\right|. \tag{B.3}$$

**Lemma 3.** Let $y=\frac{ph}{a}$ and the coefficient $\beta_1$ and $\beta_2$ given in Eq. (14) can be written as a function of $y$. Then one has:

$$\text{(i)}\ \frac{1}{12}<\beta_1<\frac{1}{6}, y<0; \beta_1=\frac{1}{12}, y=0; 0<\beta_1<\frac{1}{12}, y>0, \tag{B.4}$$

$$\text{(i)}\ 0<\beta_2<\frac{1}{12}, y<0; \beta_2=\frac{1}{12}, y=0; \frac{1}{12}<\beta_2<\frac{1}{6}, y>0, \tag{B.5}$$

**Proof:** (i) Using L'Hospital's rule, one can have the following limits:

$$\lim_{y\to\pm 0}\beta_1(y)=\frac{1}{12}, \lim_{y\to+\infty}\beta_1(y)=0, \lim_{y\to-\infty}\beta_1(y)=\frac{1}{6}. \tag{B.6}$$

The coefficient $\beta_1(y)$ is continuous on $(-\infty,0)$ and $(0,+\infty)$. The derivative $\beta_1'(y)$ can be shown as negative on $(-\infty,0)$ and $(0,+\infty)$ with simple calculus. Therefore, $\beta_1(y)$ is strictly decreasing on $(-\infty,0)$ and $(0,+\infty)$. This completes the proof of Eq. (B.4).

(ii) The proof of Eq. (B.5) is analogous.

**Lemma 4.** With given $\beta_1$ and $\beta_2$ given in Eq. (14), the boundary entries in $\mathbf{A}$ holds:

$$\text{(i)}\left|\left(\frac{1}{2}-\beta_1\right)\frac{h}{a}+\frac{1}{6}\frac{ph^2}{a^2}+\beta_1\frac{p^2h^3}{2a^3}\right| > \left|\beta_1\frac{h}{a}\right|, \tag{B.7}$$

$$\text{(ii)}\left|\left(\frac{1}{2}-\beta_2\right)\frac{h}{a}-\frac{1}{6}\frac{ph^2}{a^2}+\beta_2\frac{p^2h^3}{2a^3}\right| > \left|\beta_2\frac{h}{a}\right|. \tag{B.8}$$

**Proof.** (i) Let $y=\frac{ph}{a}$, the proof of Eq. (B.7) is equivalent to prove the following equation:

$$\left|\frac{\left(\frac{1}{2}-\beta_1\right)+\frac{1}{6}y+\frac{1}{2}\beta_1 y^2}{\beta_1}\right| = \left|\frac{1}{2\beta_1}-1+\frac{y}{6\beta_1}+\frac{y^2}{2}\right| > 1. \tag{B.9}$$

First, with $p=0$ and $\beta_1=\frac{1}{12}$, we can see that:

$$\left|\frac{1}{2\beta_1}-1+\frac{y}{6\beta_1}+\frac{y^2}{2}\right| > \left|\frac{1}{2\beta_1}-1\right| = 5 > 1. \tag{B.10}$$

For $p>0$, this gives $y>0$ and $0<\beta_1<\frac{1}{12}$. Then, we have:

$$\left|\frac{1}{2\beta_1}-1+\frac{y}{6\beta_1}+\frac{y^2}{2}\right|>\left|5+\frac{y}{6\beta_1}+\frac{y^2}{2}\right|>1. \qquad \text{(B.11)}$$

For $p$ <0, this gives $y$ < 0 and $\frac{1}{12}<\beta_1<\frac{1}{6}$. Then, we have:

$$\frac{3}{2}<\frac{1}{2\beta_1}-1-\frac{1}{72\beta_1^2}<3, \qquad \text{(B.12)}$$

and

$$\left|\frac{1}{2\beta_1}-1+\frac{y}{6\beta_1}+\frac{y^2}{2}\right|=\left|\frac{1}{2\beta_1}-1-\frac{1}{72\beta_1^2}+\left(\frac{y}{\sqrt{2}}+\frac{1}{6\sqrt{2}\beta_1}\right)^2\right|>\left|\frac{3}{2}+\left(\frac{y}{\sqrt{2}}+\frac{1}{6\sqrt{2}\beta_1}\right)^2\right|>1. \qquad \text{(B.13)}$$

This completes the proof.

(ii) The deduction for Eq. (B.8) is analogous.

**References**

[1] J. Isenberg, C. Gutfinger, Heat transfer to a draining film, Int. J. Heat Mass Transf. 16 (1973) 505–512.

[2] M. Sheikholeslami, R. Ellahi, M. Hassan, S. Soleimani, A study of natural convection heat transfer in a nanofluid filled enclosure with elliptic inner cylinder, Int. J. Numer. Methods Heat Fluid Flow. 24 (2014) 1906–1927.

[3] E.J.M. Veling, Radial transport in a porous medium with Dirichlet, Neumann and Robin-type inhomogeneous boundary values and general initial data: Analytical solution and evaluation, J. Eng. Math. 75 (2012) 173–189.

[4] J.-Y. Parlange, Water transport in soils., Annu. Rev. Fluid Mech. Vol. 12. 12 (1980) 77–102.

[5] R.S. Hirsh, Higher order accurate difference solutions of fluid mechanics problems by a compact differencing technique, J. Comput. Phys. 19 (1975) 90–109.

[6] M. Ciment, S.H. Leventhal, B.C. Weinberg, The operator compact implicit method for parabolic equations, J. Comput. Phys. 28 (1978) 135–166.

[7] B.J. Noye, H.H. Tan, A third-order semi-implicit finite difference method for solving the one-dimensional convection-diffusion equation, Int. J. Numer. Methods Eng. 26 (1988) 1615–1629.

[8] M.M. Chawla, M.A. Al-Zanaidi, D.J. Evans, Generalized trapezoidal formulas for parabolic equations, Int. J. Comput. Math. 70 (1999) 429–443.

[9] S. Karaa, J. Zhang, High order ADI method for solving unsteady convection – diffusion problems, J. Comput. Phys. 198 (2004) 1–9.

[10] M. Dehghan, Weighted finite difference techniques for the one-dimensional advection-diffusion equation, Appl. Math. Comput. 147 (2004) 307–319.

[11] A. Mohebbi, M. Dehghan, High-order compact solution of the one-dimensional heat and advection–diffusion equations, Appl. Math. Model. 34 (2010) 3071–3084.

[12] H. Ding, Y. Zhang, A new difference scheme with high accuracy and absolute stability for solving convection-diffusion equations, J. Comput. Appl. Math. 230 (2009) 600–606.

[13] Z.F. Tian, P.X. Yu, A high-order exponential scheme for solving 1D unsteady

convectiondiffusion equations, J. Comput. Appl. Math. 235 (2011) 2477–2491.
[14] Y. Ge, F. Zhao, J. Wei, A High Order Compact ADI Method for Solving 3D Unsteady Convection Diffusion Problems, Appl. Computat. Math., 7 (2018) 1–10.
[15] M.M. Chawla, M.A. Al-Zanaidi, M.G. Al-Aslab, Extended one-step time-integration schemes for convection-diffusion equations, Comput. Math. with Appl. 39 (2000) 71–84.
[16] W.F. Spotz, G.F. Carey, Extension of high-order compact schemes to time-dependent problems, Numer. Methods Partial Differ. Equ. 17 (2001) 657–672.
[17] J. Zhao, W. Dai, T. Niu, Fourth-order compact schemes of a heat conduction problem with Neumann boundary conditions, Numer. Methods Partial Differ. Equ. 23 (2007) 949–959.
[18] H.H. Cao, L. Bin Liu, Y. Zhang, S.M. Fu, A fourth-order method of the convection-diffusion equations with Neumann boundary conditions, Appl. Math. Comput. 217 (2011) 9133–9141.
[19] D. You, A high-order Padé ADI method for unsteady convection-diffusion equations, J. Comput. Phys. 214 (2006) 1–11.
[20] Z.F. Tian, Y.B. Ge, A fourth-order compact ADI method for solving two-dimensional unsteady convection-diffusion problems, J. Comput. Appl. Math. 198 (2007) 268–286.
[21] Y. Ge, Z.F. Tian, J. Zhang, An exponential high-order compact ADI method for 3D unsteady convection-diffusion problems, Numer. Methods Partial Differ. Equ. 29 (2013) 186–205.
[22] B.J. Noye, A new third-order finite-difference method for transient one-dimensional advection—diffusion, Commun. Appl. Numer. Methods. 6 (1990) 279–288.
[23] X.F. Feng, Z.F. Tian, Alternating group explicit method with exponential-type for the diffusion–convection equation, Int. J. Comput. Math. 83 (2006) 765–775.
[24] R.S. Varga, Matrix Iterative Analysis, USSR Comput. Math. Math. Phys. 5 (1999).